\documentclass[12pt]{emulateapj}

\usepackage{graphicx}

\newcommand{\hi}{{\sc Hi} }
\newcommand{\hip}{{\sc Hi}}

\newcommand{\AaA}{A\&A}
\newcommand{\ApJS}{ApJS}
\newcommand{\AJ}{AJ}
\newcommand{\MNRAS}{MNRAS}
\newcommand{\ApJ}{ApJ}
\newcommand{\Natur}{Nature}

\begin{document}

\title{First detection of dust emission in a High-Velocity Cloud}

\author{M.-A. Miville-Desch\^enes\altaffilmark{1} and F. Boulanger}
\affil{Institut d'Astrophysique Spatiale, Universit\'e Paris-XI, 91405, Orsay Cedex, France}
\and

\author{W. T. Reach \& A. Noriega-Crespo}
\affil{Spitzer Science Center, California Institute of Technology, MS 100-22, Pasadena, CA 91125, USA}

\altaffiltext{1}{Senior Research Associate, Canadian Institute for Theoretical Astrophysics, University of Toronto, 
60 St-George st, Toronto, Canada}

\begin{abstract}
By comparing sensitive Spitzer Space Telescope (SST) infrared and Green Bank Telescope 21 cm observations, 
we are able to report the first detection of dust emission in Complex C, the largest 
High Velocity Cloud in the sky. 
Dust in the region of Complex C studied here has a colder temperature 
($T=10.7^{+0.9}_{-0.8}$~K - 1$\sigma$) 
than the local interstellar medium ($T=17.5$~K), in accordance with its great distance from the Galactic plane.
Based on the metallicity measurements and assuming diffuse Galactic interstellar medium dust properties 
and dust/metals ratio, this detection could imply gas column densities more than 5 times higher 
than observed in \hip. We suggest that the dust emission detected here comes from small molecular
clumps, spatially correlated with the \hi but with a low surface filling factor. 
Our findings imply that the HVCs' mass would be much larger than inferred from \hi observations and that
most of the gas falling on the Milky Way would be in cold and dense clumps rather than in a diffuse phase.
\end{abstract}


\section{Introduction}
Since their discovery in HI observations \citep{muller63}, High Velocity Clouds (HVCs) 
have been the target of numerous studies (see \citet{van_woerden2004} for a thorough review)
but nevertheless remain puzzling. It is now widely considered that many of them 
might be infalling clouds fueling the Galaxy with low metallicity gas. 
This hypothesis received strong support from ultraviolet observations showing that HVCs have a subsolar 
metallicity \citep{wakker99} and a D/H ratio compatible with primordial abundances of deuterium \citep{sembach2004}. 

The present study focuses on the Spitzer Extra-Galactic First Look Survey (XFLS) field,
a diffuse HI area ($N_{HI} \sim 2\times10^{20}$ cm$^{-2}$) at high Galactic latitude ($l=88^\circ.3$, $b=34^\circ.9$)
programmed to study galaxy populations and clustering \citep{fang2004}.
The XFLS field is located on the edge of Complex C, a high-velocity gas structure spanning more than 
1500 square degrees on the sky. Complex C has a subsolar metallicity ($\sim 0.1-0.3$) \citep{tripp2003}, 
it contains ionised gas observed in $H\alpha$ emission \citep{tufte98a} 
and O VI absorption \citep{sembach2003} and is located at a distance greater than 5 kpc 
from the sun \citep{wakker2001}. 
Dust emission has been unsuccesfully looked for in HVCs using IRAS data \citep{wakker86}.
Upper limits were found to be compatible with their low \hi column densities, 
low metallicity (and therefore low dust/gas ratio) and their large distance (i.e. fainter radiation field). 
In this study we revisit the dust/gas correlation in HVCs using new infrared and \hi data.

\section{Data}

\label{section_data}

\begin{figure}
\includegraphics[width=\linewidth, draft=false]{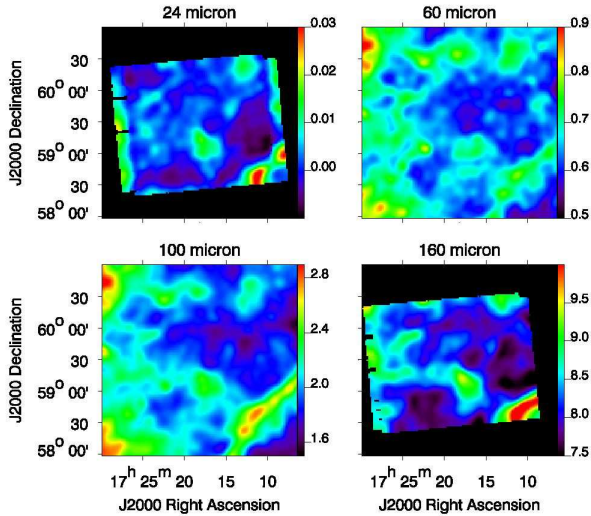}
\caption{\label{fig_infraredmaps} Infrared maps of the XFLS field at 24 and 160~micron from Spitzer
and 60 and 100 micron from IRIS. All maps were convolved 
by a Gaussian beam to bring them to the GBT 9.2' FWHM resolution.}
\end{figure}

In this study we used Spitzer Space Telescope (SST) MIPS 24 and 160~$\mu$m data of the XFLS. 
The field was covered by MIPS in 8 scan maps that were combined and destriped, 
then regridded to 16$''$ pixels and median-smoothed. 
We complemented the SST data with improved IRAS 60 and 100~$\mu$m maps (IRIS - \citet{miville-deschenes2005}).
The FWHM angular resolution of the IRIS maps is $\sim 4'$ with a pixel size of 1'.5.
The calibration uncertainties on diffuse emission of the SST and IRIS maps is 10~\%.
Bright point sources were removed from all infrared maps (SST and IRIS)
before convolution by a Gaussian beam to bring all of them to the Green Bank Telescope (GBT) 9'.2 FWHM resolution.
Fig.~\ref{fig_infraredmaps} shows the convolved maps.

The \hi (21 cm) data we used in this study were obtained with the GBT,
previously published and kindly made available by \citet{lockman2005}.
These observations cover a $3^\circ\times 3^\circ$ area centered at (J2000) 
$\alpha=17^h18^m$, $\delta=+59^\circ30'$, which encloses the SST observations. 
The GBT data were corrected for stray-radiation contamination.
The sampling was done at 1'5, the FWHM beam is 9.2', the channel width is $dv=$0.51 km s$^{-1}$ 
and the noise level in a single channel is 0.05 K. 

\begin{figure}
\includegraphics[width=\linewidth, draft=false]{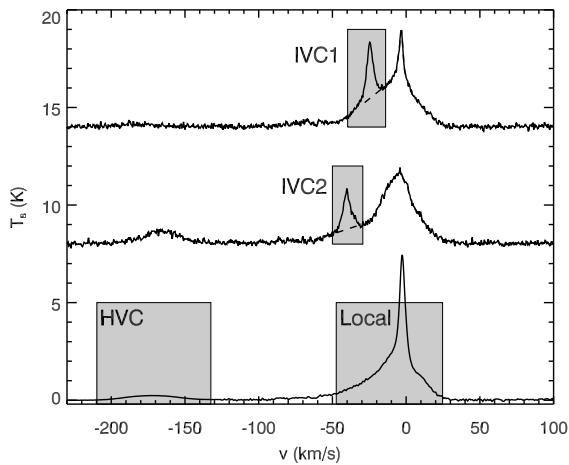}
\caption{\label{fig_spectreHI} Decomposition of the 21 cm data. The bottom curve shows the
21 cm spectrum averaged over the whole field. The middle and top curves (shifted for clarity) are representative
21 cm spectra where the IVC1 (top) and IVC2 (middle) components are observed. The shaded rectangles
indicate the velocity range used to extract the \hi components.}
\end{figure}

\section{HI decomposition}

Fig.~\ref{fig_spectreHI} presents the average 21 cm spectrum of the field and two typical lines-of-sight.
In addition to Complex C, the XFLS 21 cm data revealed two Intermediate-Velocity Clouds (IVC1 and IVC2)
and a local component typical of high-latitude \hi emission with a narrow cold component superimposed on
a broad warm component \citep{lockman2005}. 
In this section we estimate the column density of each of these \hi components.
The HVC is well isolated in velocity space and its integrated emission was computed 
by adding all the channels where $-210 < v < -132$ km s$^{-1}$.
On the other hand the two IVCs and the local component overlap in velocity.
The integrated emission of the two IVC components were computed by
adding all the channels between $-40 < v < -14$ km s$^{-1}$ for IVC1 
and $-50 < v < -29$ km s$^{-1}$ for IVC2 from which we subtracted the local component 
contribution estimated by interpolating a baseline on each spectrum (see Fig.~\ref{fig_spectreHI}). 
The local component's integrated emission was computed by adding all channels between  
$-48 < v < +25$ km s$^{-1}$
from which the two IVC contributions were removed. There is still contribution
from the local components in a few channels at $v < -48$~km s$^{-1}$ 
but they were not added, as the emission is very diffuse and its structure within the noise.

\begin{figure}
\includegraphics[width=\linewidth, draft=false]{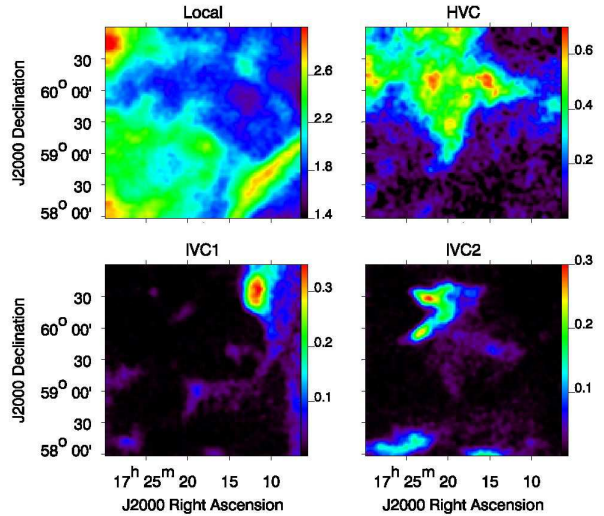}
\caption{\label{fig_himaps} \hi column density of the four \hi components (Local, HVC, IVC1 and IVC2). 
All maps are in units of $10^{20}$~cm$^{-2}$. An opacity correction was applied to convert the
21 cm integrated emission to \hi column density - see text for details.}
\end{figure}

In the optically-thin case 
(i.e. when the 21 cm brightness temperature is much smaller than the \hi kinetic temperature), 
the 21 cm integrated emission is proportional to the \hi column density:
$N_{HI}(x,y) = 1.823 \times 10^{18} \sum_v T_B(x,y,v) dv$. 
The local component is dominated by a broad (i.e. warm and optically-thin) component but a fraction of the emission comes
from cold gas (narrow features) that needs to be corrected for opacity to properly estimate the column density.
To do so we estimated the fraction of warm and cold gas on each line-of-sight
using a Gaussian decomposition of each spectrum. For each Gaussian component 
we estimated its opacity correction, considering that 
thermal and turbulence broadening equally contribute to the velocity dispersion $\sigma_v$ 
(i.e. spin temperature is $T_s = 121 \sigma_v^2 /2$), instead of the usual isothermal
assumption (\citet{lockman2005} used $T_s=80$~K). This allowed us to compute a larger correction for 
the narrow than for the broad components.
Overall the opacity correction for the local component is lower than 6~\%. 
It reaches its maximum value in the bright filament (south-western part of the field,
see Fig.~\ref{fig_himaps}) where our estimate of the \hi kinetic temperature gets as low as 50~K. 
The opacity correction for the two IVCs and the HVC is less than 1\% in accordance with their
broad lines ( FWHM$>8$~km~s$^{-1}$).
Fig.~\ref{fig_himaps} presents the \hi column density maps of the four \hi components.

\section{Infrared-HI correlation}

The main goal of this study was to separate and determine the infrared emission of the
HI components identified with the 21~cm observations. 
To do so we considered that the infrared map $I_\lambda(x,y)$ at wavelength $\lambda$ 
can be represented by the following model
\begin{equation}
\label{eq_regress}
I_\lambda(x,y) = \sum_i \alpha^{i}_\lambda N_{HI}^{i}(x,y) + C_\lambda(x,y)
\end{equation}
where $N_{HI}^i(x,y)$ is the column density of the $i^{th}$ HI component, 
$\alpha^i_\lambda$ is the infrared-\hi correlation coefficient of component $i$ 
at wavelength $\lambda$ and $C_\lambda(x,y)$ is a residual term.

\begin{figure}
\includegraphics[width=\linewidth, draft=false]{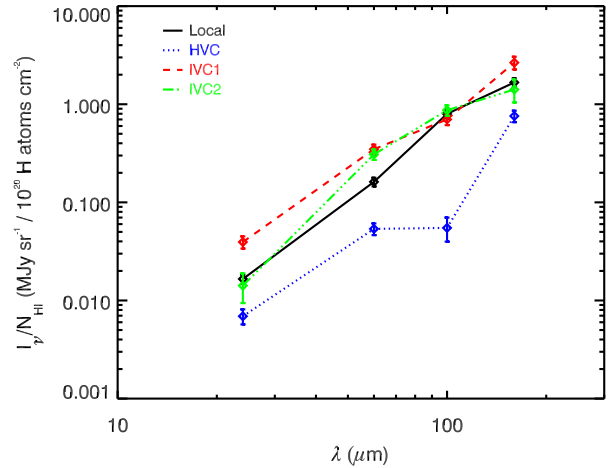}
\caption{\label{fig_sed} \hi-infrared correlation coefficients for the Local, HVC, IVC1 and IVC2 
components shown in Fig.~\ref{fig_himaps}. See Table~\ref{table_ir} for details.}
\end{figure}

To estimate the correlation coefficients $\alpha_\lambda^i$ we performed a chi-square minimisation 
on the \hi and infrared data based on the model given by Eq.~\ref{eq_regress}.
The infrared-\hi correlation coefficients
are collected in Table~\ref{table_ir} and shown in Fig.~\ref{fig_sed}.
The standard deviation of the residual maps $C_\lambda(x,y)$ are also given
in Table~\ref{table_ir}.

The uncertainties on $\alpha_\lambda^i$ given in Table~\ref{table_ir} include calibration and
statistical uncertainties, taking into account the data sampling.
To estimate the impact of our opacity correction we made the same analysis
with optically-thin \hi maps and obtained the same results within the uncertainties.

The infrared-\hi correlation presented here shows that the model used (Eq.~\ref{eq_regress}) 
for the infrared emission is satisfying.  
At each wavelength the standard deviation of the residual term $C_\lambda(x,y)$ is 
in good concordance with  the expected \citep{lagache2003} and measured \citep{miville-deschenes2002} 
levels of the Cosmic Infrared Background (CIB) fluctuations (see Table~\ref{table_ir}),
except at 160~$\mu$m where we report a significant residual excess. As the residues at 60 and 100~$\mu$m
are in accordance with the CIB level, this analysis also leads us to conclude that dust 
associated with the Warm Ionised Medium (WIM) \citep{lagache99} 
does not significantly bias our decomposition of the infrared emission in this field.

The striking result of this analysis is certainly the significant detection 
of the infrared emission associated with an HVC at all wavelengths. Our result at 100~$\mu$m is
compatible with the upper limit given by \citet{lockman2005}.

\section{Dust properties of Complex C}

Using the infrared-\hi correlation coefficients we now want to put contrains on the properties of the dust,
especially in the HVC.

We assume that the emission from big grains at thermal equilibrium with the radiation field is a modified black body :
\begin{equation}
I_\lambda = \tau_\lambda B_\nu(T_{BG})
\end{equation}
where $B_\nu$ is the Planck function, $T_{BG}$ is the big grains' equilibrium temperature and $\tau_\lambda$ the opacity, 
often expressed as $\tau_\lambda = N_{HI}\epsilon_\lambda$ where $\epsilon_\lambda$ is the dust emissivity per H atom.
The emissivity is generally described as a power law : $\epsilon_\lambda = \epsilon_0 \lambda^{-\beta}$
where $\beta$ is the emissivity spectral index.
Based on FIRAS data, \citet{boulanger96} give the following expression for the big grains' emissivity law in 
the local interstellar medium:
\begin{equation}
\epsilon_\lambda = 1.0 \times 10^{-25}(\lambda/250 \, \mu m)^{-2}\;\; \mbox{cm$^2$}.
\end{equation}

In this scheme, and at wavelengths where the emission is dominated by big grains, the infrared-\hi correlation
coefficients can be expressed as $\alpha_\lambda = \epsilon_\lambda B_\nu(T_{BG})$.
Assuming that Very Small Grains (VSGs) do not contribute significantly in the far-infrared and
taking $\beta=2$ \citep{draine84,boulanger96}, we use 
the 100 and 160~$\mu$m infrared-\hi correlation coefficients 
to estimate the big grains' temperature $T_{BG}^i$ of each \hi component 
\begin{equation}
\frac{\alpha_{100}^i}{\alpha_{160}^i} = \frac{ B_\nu(100\mu m, T_{BG}^i) 160^2}{ B_\nu(160\mu m, T_{BG}^i) 100^2}
\end{equation}
and the dust emissivities at 160~$\mu$m : $\epsilon_{160}^i = \alpha_{160}^i/B_\nu(160 \mu m, T_{BG}^i)$.
The big grain temperature and the 160~$\mu$m emissivity of each \hi component are given
in Table~\ref{table_dust}. In this table are also given the corresponding 1 and 3$\sigma$ ranges
on $T_{BG}^i$ and $\epsilon^i_{160}$ computed using 1 and 3$\sigma$ 
uncertainties on $\alpha_{100}^i$ and $\alpha_{160}^i$.

\begin{table*}
\begin{center}
\caption{\label{table_ir} Infrared-\hi correlation}
\begin{tabular}{ccccccc} \hline
$\lambda$ & $\alpha_\lambda^{Local}$& $\alpha_\lambda^{HVC}$  & $\alpha_\lambda^{IVC1}$ & $\alpha_\lambda^{IVC2}$ & $\sigma$(Residue) & CIB estimate\\ 
($\mu$m) & & & & & (MJy sr$^{-1}$) & (MJy sr$^{-1}$)\\\hline
24 & 0.017$\pm$0.002 & 0.007$\pm$0.002 & 0.039$\pm$0.006 & 0.014$\pm$0.005 & 0.004 & 0.004\\
60 & 0.16$\pm$0.02 & 0.05$\pm$0.01 & 0.35$\pm$0.04 & 0.31$\pm$0.04 & 0.03 & 0.02\\
100 & 0.80$\pm$0.08 & 0.055$\pm$0.015 & 0.70$\pm$0.09 & 0.9$\pm$0.1 & 0.09 & 0.07 \\
160 &  1.7$\pm$0.2 &  0.8$\pm$0.1 & 2.7$\pm$0.4 & 1.4$\pm$0.4 & 0.30 & 0.11 \\\hline
\end{tabular}
\end{center}
Infrared-\hi correlation coefficients ($\alpha_\lambda^i$) given in MJy sr$^{-1}$ ($10^{20}$ H atoms)$^{-1}$ cm$^2$ for the four
\hi components (Local, HVC, IVC1 and IVC2) at 24, 60, 100 and 160~$\mu$m. The uncertainties on each
$e_\lambda^i$ is the 1$\sigma$ uncertainty taking into account the statistical variance
and instrumental uncertainties.
Column 6 gives the standard deviation of the residual map and Column 7 is an estimate of the Cosmic Infrared Background (CIB) 
fluctuations based on the model of \citet{lagache2003}.
\end{table*}

\begin{table}
\begin{center}
\caption{\label{table_dust} Dust parameters}
\begin{tabular}{lcccccc}\hline
 & $T_{BG}$ & $1\sigma$ & $3\sigma$ & $\epsilon_{160}$ &  $1\sigma$ & $3\sigma$\\ \hline
Local & 17.4 & 16.3-18.8 & 14.3-22.6 & 3.0 &  1.9-4.9 & 0.60-12.5 \\
HVC & 10.7 & 9.9-11.6 &  7.7-13.6 & 36.0 & 16.8-83.7 & 3.8-1442.8 \\
IVC1 & 14.6 & 13.6-15.8 & 11.8-19.4 & 13.2 & 7.0-24.1 & 1.6-79.6 \\
IVC2 & 19.4 & 17.1-23.1 & 14.1-107.4 & 1.5 & 0.5-3.6 & 0.003-15.5\\ \hline
\end{tabular}
\end{center}
Big Grain temperature ($T_{BG}$, given in Kelvin) and dust emissivity at 160~$\mu$m  
($\epsilon_{160} = \tau_{160}/N_{HI}$ given in $10^{-25}$ cm$^2$ per H atom) and their
corresponding 1 and 3~$\sigma$ ranges for each \hi component.
The values for the local interstellar medium deduced from the COBE data \citep{boulanger96} are
$T_{BG}=17.5$~K and $\epsilon_{160} = 2.4\times10^{-25}$~cm$^2$.
\end{table}

\section{Discussion}

The values of $T_{BG}^{local}=17.4^{+1.4}_{-1.1}$ and 
$\epsilon_{160}^{local}=3.0^{+1.9}_{-1.1}\times 10^{-25}$~cm$^2$ per H atom found for the
local component are in very good agreement (within 1$\sigma$
uncertainties) with the average values found for the Solar neighborhood 
by \citet{boulanger96} using COBE data : $T_{BG}^{soln}=17.5$~K and 
$\epsilon_{160}^{soln}=2.4 \times 10^{-25}$ cm$^2$ per H atom. 
This is an important validation of the photometric accuracy of the SST-MIPS data.
On the other hand dust in the HVC is found to be significantly colder than in the local interstellar medium, 
even at the 3$\sigma$ limit ($T_{BG}^{HVC} = 10.7_{-3.0}^{+2.9}$),  
which is consistant with a lower radiation field than in the Solar neighborhood 
due to its distance to the Galaxy \citep{wakker86}.
In addition we report an emissivity $\epsilon_{160}$ higher in the HVC than in the local ISM 
(see Table~\ref{table_dust}).
An eventual contribution of VSGs to the far-infrared emission 
would have the effect of lowering $T_{BG}$ and increasing $\epsilon_{160}$ even more.

For dust properties typical of the local interstellar medium, 
$\epsilon_\lambda$ should be proportional to the dust-to-gas mass ratio which, for standard metal 
depletion on grains, would scale with the metallicity. 
The fact that $\epsilon^{HVC}_{160}/\epsilon^{soln}_{160} > 1.6$ (3$\sigma$) is greater than the metallicity 
in Complex C ($ Z_{HVC}/Z_\odot = 0.2\pm0.1$ according to \citet{tripp2003}) could indicate a dust 
(and therefore gas) column density in Complex C more than 5 times larger than given by the \hi emission. 

Could it be dust associated with ionised gas ?
H$\alpha$ and other ionised atomic lines have been detected in several HVCs. It is expected
to come from a photoionized layer at the surface of the clouds which plunge into the Galactic halo.
The column density of such a layer is $2-4 \times 10^{19}$ cm$^{-2}$ in Complex C \citep{wakker99} 
which is similar to the \hi column density in the HVC seen in XFLS and therefore not enough 
to explain the high infrared emissivity measured here. In addition only the ionised gas spatially
correlated with the \hi would contribute to the infrared emissivity.
Could it be dust associated with diffuse H$_2$ ?
Two detections of H$_2$ in absorption with FUSE have been reported but the column densities are much lower 
than those of \hi \citep{richter2001}. Other attemps have been unsuccesful; 
CO was not found in emission \citep{wakker97} nor HCO$^+$ in absorption \citep{combes2000}. 
Furthermore abundance measurements in HVCs show no sign of iron depletion \citep{murphy2000,tripp2003}
which has been used as a strong argument in favor of very low dust in HVCs.

To reconcile our findings with previous work we suggest that HVCs could be multiphase gas
including small and dense molecular clumps with a low surface filling factor (and therefore 
undetected with pencil beam absorption observations) bathing in a diffuse phase seen in absorption.
The dust emission reported here would mainly originate in these clumps, spatially correlated with \hip, 
which would have a larger iron depletion than in the diffuse phase of HVCs.
Evidence of the existence of such small molecular clumps was found in cirrus 
clouds \citep{heithausen2004} and in the Magellanic Stream \citep{sembach2001}. 
Additional observations of HVCs with the Spitzer Space Telescope are needed to confirm our result.

If our result reveals a general property of HVCs, it would mean that their mass 
is much larger than what can be inferred from \hi observations and that most
of the gas is in a dense phase. An infall of 1 M$_\odot$/year is needed to explain 
the star formation rate, the distribution of stellar metallicities with age 
and the presence of a bar in the Milky Way and in other galaxies.
The current idea is that this infalling gas is diffuse (ionised and neutral)
resulting from the cooling of very hot halo gas. 
Our result changes this picture by showing that most of the infalling gas is in dense and cold clumps,
leaving open the question of the origin and stability of such structures.

HVCs typically contribute to $<10\%$ of the \hi column density at high latitude, but
with its temperature lower than in the local interstellar medium, 
dust in HVCs could contribute a much larger fraction to the sub-millimeter and millimeter emission.
Observations in this wavelength range, with Herschel and Planck for instance, 
could be used as a tracer of the column density of HVCs and as a distance indicator, 
using the dust temperature to put constraints on the radiation field strength.
If the detection presented in this study is confirmed and is typical of HVCs, 
it could also have a strong impact on the analysis of observational cosmology data 
(e.g. WMAP, Planck) that do not to this day take this foreground component into account.

The authors thank the anonymous referee for a careful reading of the manuscript
and very helpful comments. M.A.M.D. acknowledges support from The Canadian Space Agency.
This research has made use of NASA's Astrophysics Data System Bibliographic Services.


\begin{thebibliography}{{Miville-Deschenes} \& {Lagache}(2005)}

\bibitem[{Boulanger} et~al.(1996)]{boulanger96}
{Boulanger}, F., et al.\
\newblock 1996, \AaA, 312, 256.

\bibitem[{Combes} \& {Charmandaris}(2000)]{combes2000}
{Combes}, F. \& {Charmandaris}, V.
\newblock 2000, \AaA, 357, 75.

\bibitem[{Draine} \& {Lee}(1984)]{draine84}
{Draine}, B.~T. \& {Lee}, H.~M.
\newblock 1984, ApJ, 285, 89

\bibitem[{Fang} et~al.(2004)]{fang2004}
{Fang}, F. et~al.
\newblock 2004, \ApJS, 154, 35.

\bibitem[Heithausen(2004)]{heithausen2004} Heithausen, A.\ 2004, 
\apjl, 606, L13 

\bibitem[{Lagache} et~al.(1999)]{lagache99}
{Lagache}, G., {Abergel}, A., {Boulanger}, F., {Desert}, F.~X., {Puget}, J.~L.
\newblock 1999, \AaA, 344, 322

\bibitem[{Lagache} et~al.(2003)]{lagache2003}
{Lagache}, G., {Dole}, H. \& {Puget}, J.~L.
\newblock 2003, \MNRAS, 338, 555.

\bibitem[{Lockman} \& {Condon}(2005)]{lockman2005}
{Lockman}, F.~J. \& {Condon}, J.~J.
\newblock 2005, \aj, 129, 1968

\bibitem[{Miville-Deschenes} \& {Lagache}(2005)]{miville-deschenes2005}
{Miville-Deschenes}, M.-A. \& {Lagache}, G.
\newblock 2005, \ApJS, 157, 302

\bibitem[{Miville-Desch\^enes} et~al.(2002)]{miville-deschenes2002}
{Miville-Desch{\^e}nes}, M.~A., {Lagache}, G., \& {Puget}, J.-L..
\newblock 2002, A\&A, 393, 749.

\bibitem[{Muller} et~al.(1963)]{muller63}
{Muller}, C.~A., {Oort}, J.~H. \& {Raimond}, E.
\newblock 1963, Comptes rendus de l'Acad\'emie des Sciences de Paris, 257, 1661.

\bibitem[Murphy et al.(2000)]{murphy2000} Murphy, E.~M., et al.\ 
2000, \apjl, 538, L35 

\bibitem[{Richter} et~al.(2001a)]{richter2001}
{Richter}, P., {Sembach}, K.~R., {Wakker}, B.~P. \& {Savage}, B.~D.
\newblock 2001a, \ApJ, 562, L181.

\bibitem[{Richter} et~al.(2001b)]{richter2001a}
{Richter}, P., et al.\
\newblock 2001b, \ApJ, 559, 318.

\bibitem[Sembach et al.(2001)]{sembach2001} Sembach, K.~R., Howk, 
J.~C., Savage, B.~D., \& Shull, J.~M.\ 2001, \aj, 121, 992 

\bibitem[{Sembach} et~al.(2003)]{sembach2003}
{Sembach}, K.~R. et~al.
\newblock 2003, \ApJS, 146, 165.

\bibitem[{Sembach} et~al.(2004)]{sembach2004}
{Sembach}, K.~R. et~al.
\newblock 2004, \ApJS, 150, 387.

\bibitem[{Tripp} et~al.(2003)]{tripp2003}
{Tripp}, T.~M. et~al.
\newblock 2003, \AJ, 125, 3122.

\bibitem[{Tufte} et~al.(1998)]{tufte98a}
{Tufte}, S.~L., {Reynolds}, R.~J. \& {Haffner}, L.~M.
\newblock 1998, \ApJ, 504, 773.

\bibitem[{van Woerden} et~al.(2004)]{van_woerden2004}
{van Woerden}, H., {Wakker}, B.~P., {Schwarz}, U.~J. \& {De Boer}, K.~S.
\newblock 2004, {\em High Velocity Clouds}.
\newblock Kluwer Academic Publishers, Dordrecht.

\bibitem[{Wakker} \& {Boulanger}(1986)]{wakker86}
{Wakker}, B.~P. \& {Boulanger}, F.
\newblock 1986, \AaA, 170, 84.

\bibitem[{Wakker} et~al.(1997)]{wakker97}
{Wakker}, B.~P., {Murphy}, E.~M., {van Woerden}, H. \& {Dame}, T.~M.
\newblock 1997, \ApJ, 488, 216.

\bibitem[{Wakker} et~al.(1999)]{wakker99}
{Wakker}, B.~P. et~al.
\newblock 1999, \Natur, 402, 388.

\bibitem[{Wakker}(2001)]{wakker2001}
{Wakker}, B.~P.
\newblock 2001, \ApJS, 136, 463.

\end{thebibliography}
\end{document}